\documentclass{article}
\usepackage[a4paper, total={6in, 8in}]{geometry}
\usepackage[labelfont=bf]{caption}

\usepackage{amsmath,amsfonts,amssymb,amsthm,booktabs,color,epsfig,graphicx, url}

\theoremstyle{plain}

\theoremstyle{definition}

\usepackage{natbib}
\usepackage{multirow}
\usepackage{enumitem}
\usepackage{xcolor}
\newcommand{\norm}[1]{\left\lVert#1\right\rVert}

\allowdisplaybreaks
\begin{document}
\def\spacingset#1{\renewcommand{\baselinestretch}%
{#1}\small\normalsize} \spacingset{1}


  \title{An efficient approach to characterize spatio-temporal dependence in cortical surface fMRI data}
  \author{Huy Dang\\
    Dept.~of Statistics, Pennsylvania State University, USA\\
    and \\
    Marzia A. Cremona \\
    Dept.~of Operations and Decision Systems, Université Laval, Canada\\
    CHU de Québec – Université Laval Research Center, Canada\\
    and \\
    Francesca Chiaromonte\\
    Dept.~of Statistics, Pennsylvania State University, USA\\
    Inst.~of Economics and L'EMbeDS, Sant'Anna School of Advanced Studies, Italy \\
    and \\
    Nicole Lazar\\
    Dept.~of Statistics, Pennsylvania State University, USA
    }
  \maketitle

\bigskip
\begin{abstract}
\noindent

Functional magnetic resonance imaging (fMRI) is a neuroimaging technique known for its ability to capture brain activity non-invasively and at fine spatial resolution (1-3mm). 
Cortical surface fMRI (cs-fMRI) is a recent development of fMRI that restricts attention to signals exclusively from tissue types that have neuronal activities, as opposed to the whole brain.
cs-fMRI data is plagued with non-stationary spatial correlations and long temporal dependence which, if inadequately accounted for, can hinder various types of downstream statistical analyses.
We propose a fully integrated approach that captures both spatial non-stationarity and varying ranges of temporal dependence across regions of interest. More specifically, we impose non-stationary spatial priors on the latent activation fields and model temporal dependence via fractional Gaussian errors of varying Hurst parameters, which can be studied through a wavelet transformation and its coefficients' variances at different scales. We demonstrate the performance of our proposed approach via simulations and an application to a visual working memory task cs-fMRI dataset.
\end{abstract}

\noindent%
{\it Keywords:}  
cs-fMRI, spatio-temporal dependence, SPDE, wavelets, fractional Gaussian noise. 

\newpage
\spacingset{1.5} 

\section{Introduction and motivation\label{sec:1}}

\begin{figure}[!ht]
        \hspace{-0.5cm}
        \begin{tabular}{cc}
        \includegraphics[width=.48\textwidth]{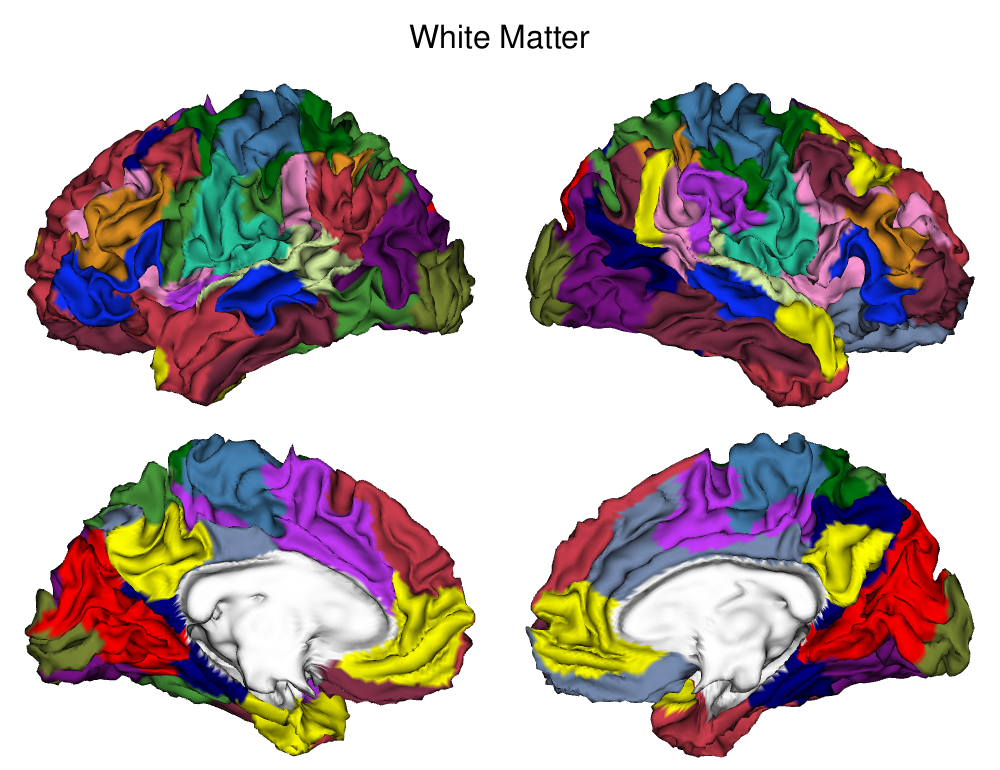}&
        \includegraphics[width=.48\textwidth]{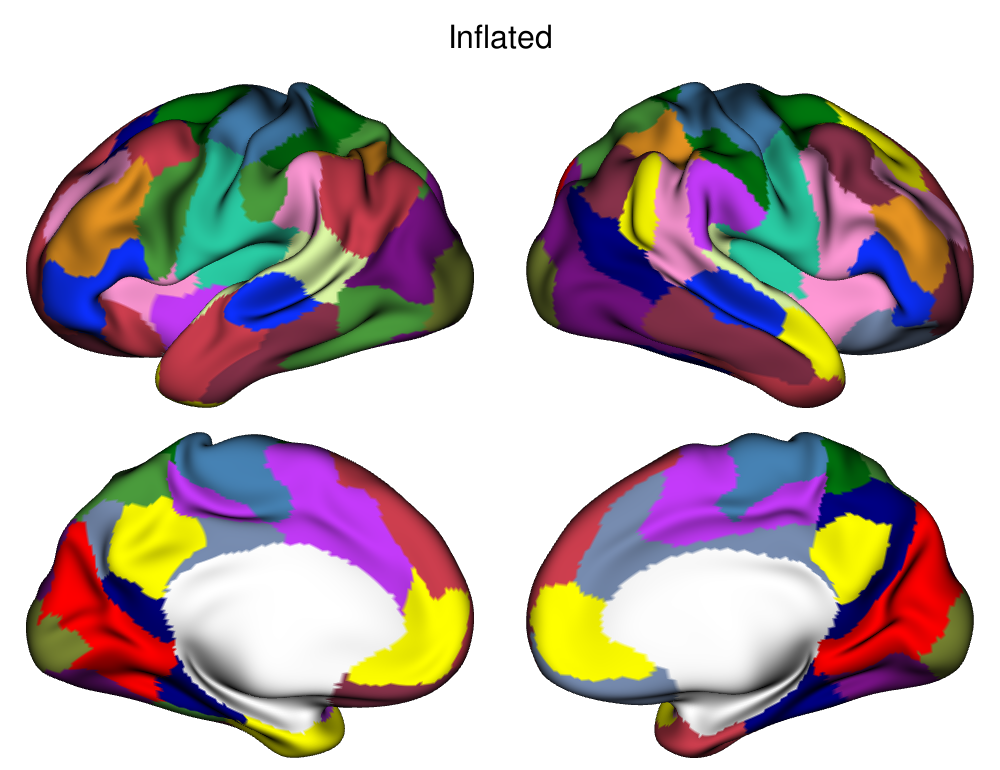}\\
        \includegraphics[width=.48\textwidth]{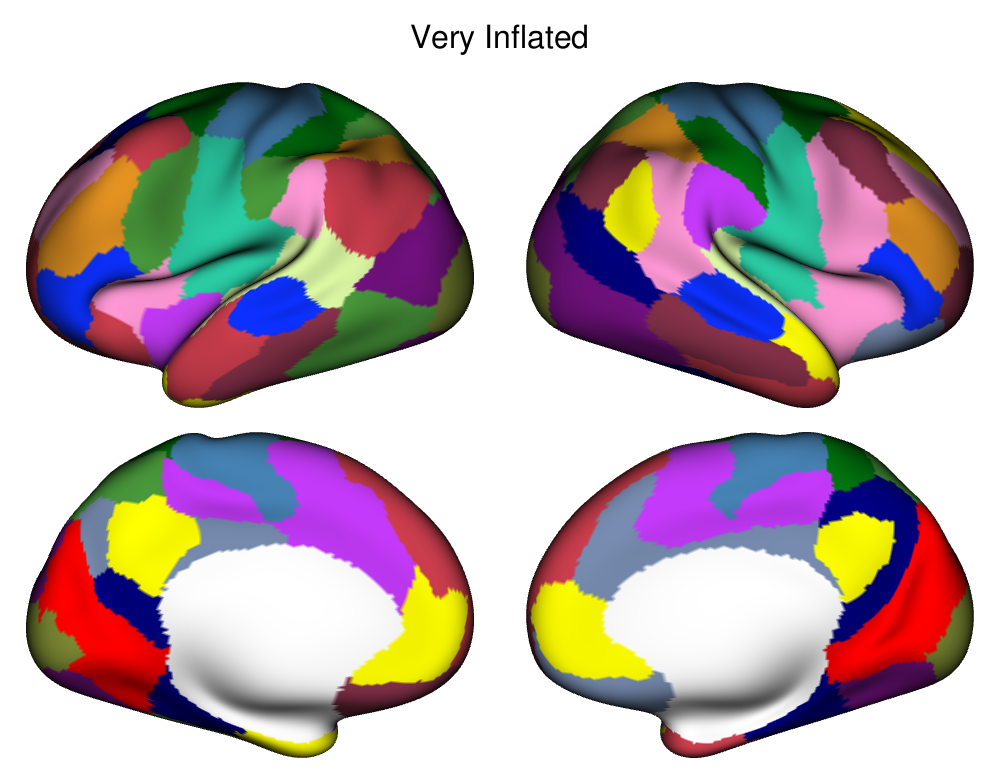}&
        \includegraphics[width=.48\textwidth]{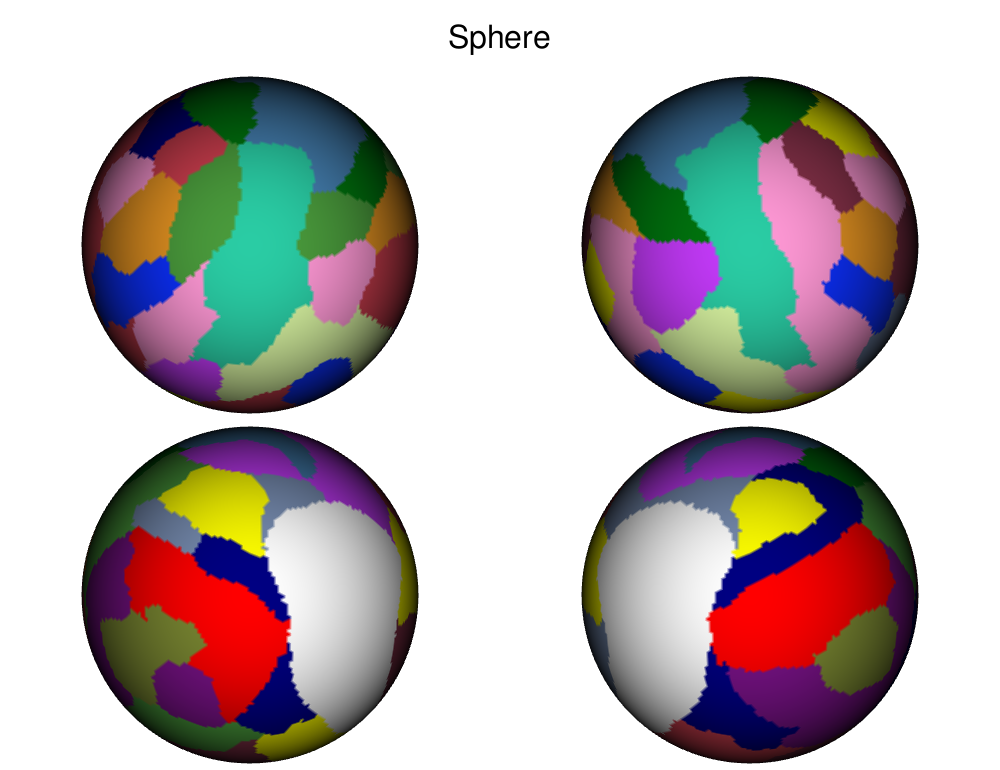}
        \end{tabular}
        \caption{
        The 2D cortical surface mesh used in cs-fMRI data, visualized with increasing degrees of inflation: white matter surface, inflated, very inflated, spherical. The mesh is colored according to a 
        parcellation of the brain into 100 regions (50 regions per hemisphere) introduced in \citet{schaefer18}. Subcortical regions are plotted in white and are not analyzed.}
        \label{fig:surface} 
    \end{figure}

Functional magnetic resonance imaging (fMRI) is a neuroimaging technique
known for its ability to non-invasively measure 
Blood Oxygen Level Dependent (BOLD) signals at fine spatial resolution (typically 2-3mm). The BOLD signal is defined as the changes in the ratio of oxygenated to deoxygenated blood, either in response to a task/stimulus in an experiment or as a consequence of spontaneous  neural metabolism. 
In fMRI literature, the BOLD signal is widely taken to be a proxy of brain activity, and 
used to capture
associations between local brain areas with 
different functions. 
Traditionally, 
fMRI data collected during a task
are represented by 4D arrays:
signals are measured at discrete locations, each representing a 
small cube in a partition of the 3D brain (these are
known as voxels), along a time course -- which produces the 4th dimension. 
This type of
``volumetric'' fMRI data
contains both signals from
tissues that have neuronal activities (such as gray matter) and noise from
tissues that do not (such as white matter and cerebral spinal fluid). 
Because of spatial contiguity between active and inactive tissues, distance-based analyses are often affected by spurious noise from the latter.
Cortical surface fMRI data (cs-fMRI) is the result of a recently developed technique that uses vertices on a 2D surface mesh, instead of traditional volumetric voxels, to represent the folded, sheet-like geometry of the cerebral cortex. 
In multi-subject studies
surface meshes from different individuals can be aligned to a common template based on cortical folding patterns and areal features \citep{glasser13,glasser16}. 
By virtue of being confined exclusively to cortical gray matter, 
cs-fMRI data is not affected by signal contamination from
inactive tissues.
Increased homogeneity 
due to the fact that data are collected only from
cortical gray matter 
thus improves distance-based analyses.
Figure~\ref{fig:surface} 
illustrates the surface mesh 
used to achieve the 2D representation of the cerebral cortex at different degrees of ``inflation''
-- ranging from ``white matter surface'' (least inflated and preserving folding patterns) to ``inflated'', ``very inflated'', and ``spherical'' (most inflated, no folding patterns). 
By inflating the surface mesh, distances between mesh vertices lose their original 3D Euclidean interpretation and become more geodesic-like. 
Readers interested in a more comprehensive review of volumetric and cs-fMRI data are referred to \citet{mejia20}. 
From 
here on, ``voxel'' refers to the smallest spatial unit in volumetric fMRI data, and ``vertex'' 
to its cs-fMRI counterpart. 

\subsection{Temporal dependence}
\label{sec:temporal_dependence}
It is well known that fMRI data exhibit
temporal correlations. 
Existing literature 
provides evidence of both short- and long-range temporal dependence that can be attributed to spontaneous, non task-related fluctuations 
in neuronal activity \citep[see e.g.][]{bullmore03, park10, li13}. 
Moreover, \citet{park10,li13} showed that the ranges of dependence 
can be affected by 
brain structures and functions. 
\begin{figure}[!tbhp]
        \centering
        \includegraphics[width=\textwidth, height = .5\textwidth]{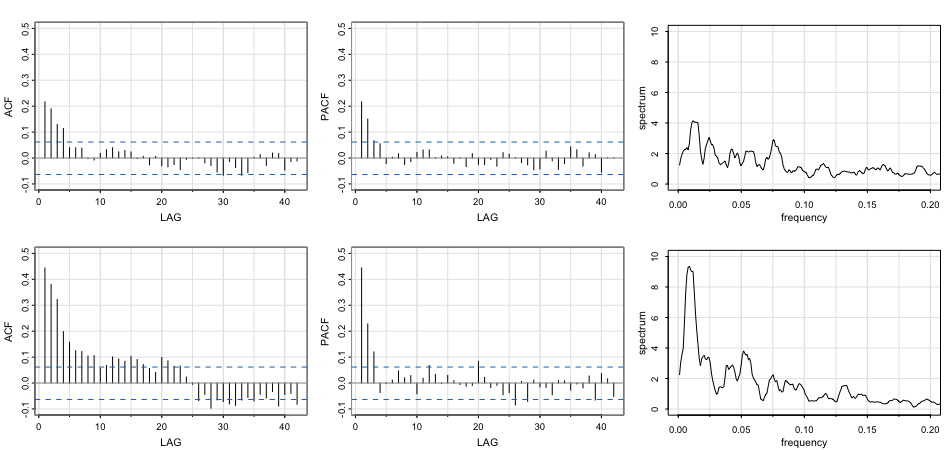} 
        \caption{Autocorrelation function (left), partial autocorrelation function (middle) and spectrum-frequency plot (right) for time series with short-range 
        (top) and long-range dependence (bottom).}
        \label{fig:temporal_dependence}
\end{figure}
As a real data illustration,
Figure \ref{fig:temporal_dependence} 
shows the autocorrelation function (ACF), the partial autocorrelation function (PACF) and the spectrum against frequency for two resting state BOLD time series. 
These are taken at different brain locations from an individual in the Human Connectome Project (HCP) healthy young adults data set \citep{vanessen13}. 
The upper panels 
illustrate
a time series with short-range temporal dependence, 
as evidenced by the presence of only a few significant ACF and PACF coefficients at small lags, 
while the 
bottom panels 
illustrate a time series with long-range temporal dependence,
with a slowly decaying autocorrelation. 
The latter also 
shows very large estimated spectral density at 
small frequencies, in contrast to the more evenly distributed spectral density across frequencies of the former. 
Since the spectral density is the Fourier transform of the covariance function, 
large spectral density 
estimates corresponds to large 
contributions to the covariance function's second moment, implying significant dependence at frequencies near $0$, or equivalently, at large lags. 

The vast majority of fMRI literature 
employs autoregressive models of order $p$ 
-- that we denote as $AR(p)$ -- to prewhiten the time series separately at each 
voxel, or vertex, prior to 
further analyses. 
Some recent examples include the use of $AR(1)$ in \citet{derado10}, $AR(2)$ in \citet{castruccio18}, $AR(3)$ in \citet{risk16}, and $AR(6)$ in \citet{mejia20}. 
However, ordinary $AR(p)$ models are designed to capture autocorrelations up to a finite order $p$, and therefore are not suitable for modeling long-range dependence. 
In addition, we 
show with a simple simulation that prewhitening time series with an AR process may cause loss of information and lead to biased estimates. 
Let $\boldsymbol{x} = \{x_t\} = \int_{-\infty}^{\infty} h(u)s(t-u)du$ be a stimulus $h(\cdot)$ convolved with the canonical hemodynamic response function $s(\cdot)$
observed along times, say,
$t = 1, 
\ldots, 256$ (more 
detail on this 
convolutional representation, which is typical in fMRI data, will
be provided in Section \ref{sec:modeling_approach}).
We simulate the responses $\boldsymbol{y}$ and $\boldsymbol{z}$ by adding to $\boldsymbol{x}$ a long-range dependent noise $\boldsymbol{\epsilon}$ and an auto-regressive noise $\boldsymbol{\eta}$, respectively; that is
\begin{align*}
    \boldsymbol{y} = \beta\boldsymbol{x} + \boldsymbol{\epsilon} 
    \; ,
    \;
    \boldsymbol{z} = \beta\boldsymbol{x} + \boldsymbol{\eta} \ .
\end{align*}
To simulate the long-range dependent
$\boldsymbol{\epsilon}$, we use fractional Gaussian noise with Hurst parameter $H = 0.8$. 
For the auto-regressive 
$\boldsymbol{\eta}$, we 
fix $p = 6$ and use 
AR coefficients 
estimated from actual BOLD signals
(from the same HCP individual used above, at randomly selected vertices/locations). 
In both cases, we set $\beta = 2$ and noise variance equal to $1$. 
We then carry out 
the estimation of $\beta$ 
following the procedure used in \citet{mejia20}: 
we fit $AR(6)$ models to
both $\boldsymbol{\epsilon}$ and $\boldsymbol{\eta}$,
obtain the AR coefficients in each of the two cases
and use them to pre-whiten $\boldsymbol{y}$ and $\boldsymbol{z}$, 
and finally regress the pre-whitened time series $\boldsymbol{y}_{pw}$ and $\boldsymbol{z}_{pw}$ 
on $\boldsymbol{x}$ to 
estimate $\beta$. 
\begin{table}[!b]
    \centering
    \begin{tabular}{ |c|c|c|c|c|c|c|c|c| } 
    \hline
    Dependence & Method & Min & Q1 & Median & Mean & Q3 & Max & SD \\
    \hline
    & \text{Ordinary} & 1.189 &  1.831 &  2.015 &  2.008 &  2.186 &  3.010  & 0.263\\
    \cline{2-9}
    \multirow{-2}*{\text{Long}} & \text{$AR(6)$ prewhitening} & 0.688 &  1.111 &  1.242 &  1.239 &  1.353 &  1.8285  & 0.182\\
     \hline
    & \text{Ordinary} & 0.482 & 1.770 & 2.013 & 2.009 & 2.244 & 2.997 & 0.343\\
    \cline{2-9}
    \multirow{-2}*{\text{$AR(6)$}} & \text{$AR(6)$ prewhitening} & 0.476 & 1.106 & 1.278 & 1.283 & 1.455 & 2.025 & 0.246\\
    \hline
    \end{tabular}
    \caption{Summaries 
    for $\hat{\beta}$ estimates in long- and short-range dependence scenarios, obtained with 
    $AR(6)$ pre-whitening 
    and with an ordinary regression
    without pre-whitening
    (results based on 1000 simulation runs).}
     \label{table:ar-bias}
\end{table}
Table \ref{table:ar-bias} shows 
estimation results 
in the long- and short-range dependence scenarios
with the $AR(6)$ prewhitening scheme,
as well as with an ordinary regression 
without prewhitening. 
The results, based on 1000 simulation runs, indicate that ordinary regression
(without prewhitening)
produces unbiased estimates of $\beta$. 
On the contrary, $AR(6)$ prewhitening of the time series 
produces a marked underestimation
of $\beta$ in both scenarios, 
including 
when the noise is in fact generated 
with an $AR(6)$ process. 

This simple but informative simulation exercise illustrates the problems that can be induced by
prewhitening time series with autoregressive models prior to 
further analyses. 
In order to account for temporal correlations, we strive for a more flexible modeling approach that can accommodate both short- and long-range dependence across brain regions. 
Such approach should not cause loss of information or biases, 
and should be implementable 
through a computationally efficient algorithm.
In particular, to contain computational cost, it 
should produce 
sparse
temporal correlations. 

\subsection{Spatial dependence}
The sheer size of fMRI data presents a challenge 
when modeling correlations between voxels or vertices.
In addition to sparsity in representing temporal dependences, 
sparsity 
assumptions on the spatial dependence structure are absolutely critical;
without them, most standard analyses (e.g., fitting linear models) will face an immense computational burden 
due to the need to invert 
a dense $VT \times VT$ covariance matrix, where $V$ is the number of voxels or vertices,
and $T$ is the number of time points. 
In traditional volumetric fMRI data $V \approx 120,000$, whereas in cs-fMRI $V \approx 30,000$ for each hemisphere,
and $T$ typically ranges from $200$ to $1,200$. 
In 
existing fMRI literature, the most common (and unrealistic) assumption 
is that of no spatial 
dependence,
which means fitting a model for each time series at each 
voxel/vertex independently -- an approach that is often referred to as {\em massively univariate } \citep{lazar08}. 
However, untreated spatial correlations
often 
lead to underestimated uncertainty and inflated type I error rates in hypothesis tests 
\citep{castruccio18}. 

While inverting the aforementioned covariance matrix has been and remains computationally unrealistic,
recent years have seen a rise in 
efforts to acknowledge and treat spatial dependence. 
This is typically achieved via a combination of more sophisticated statistical models, approximation techniques, and/or down-sampling of the data. 
An early example of spatial dependence treatment is the work of \citet{bowman05},
employing a two-stage hierarchical Bayesian approach.
First, standard regressions were fitted independently 
at each voxel 
to estimate activation coefficients; then, coefficients of voxels belonging to the same functional region were 
used 
to model within region correlations via a spatial autoregressive model. Computation was made feasible by subsampling the 3D volumetric data to 2D slice data, and by assuming stationarity of the covariance. 
A different approach was taken by \citet{kang13}, which applied a Fourier transform to the fMRI time series and fitted a spatio-spectral mixed-effect model, using random effects to capture spatial correlation within and between regions of interest. Although some degrees of non-stationarity was allowed in this approach, activation was restricted to regional level 
instead of individual voxels. 
More recently, \citet{castruccio18} used $L$ Gaussian processes with different Matern covariances to model local dependence within different regions, and a region-specific random effect to account for between-region dependence. While the model incorporated non-stationary spatial dependence on full 3D volumetric data, estimation at each spatial scale (local and regional) was carried out sequentially and only a
portion of the data was used at each estimation step. For example, local dependence was estimated region by region, and regional dependence was estimated using only regional averages. 
On a different front, \citet{mejia20} assumed a latent task activation model and 
represented spatial dependence via spatial priors on the activation fields. The proposed method was computationally very efficient, as estimation was carried out within the Gaussian Markov Random Fields (GMRFs) framework, and inference was made possible in continuous space thanks to an established connection between GMRFs and Gaussian Fields (GFs) with Matern covariance functions \citep{lindgren11}. Their approach, however, did not allow for non-stationarity, which is an important spatial feature of fMRI data. 


\subsection{On things to come}
To improve on the existing treatment of fMRI data, we propose a fully integrated approach that captures both spatial non-stationarity and varying ranges of temporal dependence across regions of interest, focusing in particular on the new class of 
cs-fMRI data. 
More specifically, we impose spatial priors on the latent activation fields as in \citet{mejia20}; however, our 
approach allows for non-stationarity by letting the prior hyperparameters 
be driven by local smoothness in the data.
We model temporal dependence
using fractional Gaussian
errors of varying Hurst parameters; the Hurst parameters can then be studied 
through wavelet transformation and the wavelet coefficients' variances at different scales. 
Bayesian inference is carried out 
approximating the marginal posteriors
with the Integrated Nested Laplace Approximation (INLA) approach \citep{rue09}. 

The remainder of this 
article is organized as follows. Section~2
provides the theoretical background. Section~3
details our modeling approach. 
Section~4
demonstrates the performance of our proposal through simulations and an application to real cs-fMRI data concerning a visual working memory task. Section~5
contains final remarks.

\section{Theoretical background\label{sec:2}}

\subsection{Fractional Gaussian noise and wavelet transformation}
\label{sec:dwt}
Long- and short-range dependence can be flexibly and sparsely modeled with 
fractional Gaussian noise (fGn). fGn is defined as the difference between consecutive values of fractional Brownian motion \citep{fadili02}. Given a lag $\ell > 0$, the autocovariance of a fractional Gaussian process $G$ is given by
\begin{align}
\label{fgn}
    C_G(\ell; H) = \frac{\sigma^2}{2}[(\ell + 1)^{2H} - 2\ell^{2H} + (\ell -1)^{2H}] \underset{\ell \rightarrow \infty}{\approx} \sigma^2 H(2H-1)\ell^{2H -2}
\end{align}
where $H$ is the Hurst parameter and $\sigma^2$ is the variance. 
It is easy to see that when $H = \frac{1}{2}$, the process 
corresponds to white noise. 
When $0 < H < \frac{1}{2}$, or equivalently, $-2 < 2H - 2 < -1$, the autocovariance decays exponentially fast with lag $\ell$ and the process has short-range dependence. 
On the other hand, when $ \frac{1}{2} < H < 1$, the autocovariance decays at hyperbolic rate $-1 < 2H - 2 < 0$, causing $\sum_{\ell = 0}^\infty C_G(\ell; H) = \infty$, and the process has long-range dependence. When $\sum_{\ell = 0}^\infty C_G(\ell; H) = \infty$, individual correlations at 
large lags may be small, but cumulatively, 
their effect is significant. A consequence of untreated correlations 
are 
biases in variance estimates that do not 
vanish as the sample size increases, which in turn affects confidence intervals and hypothesis tests. 

To treat time series with varying ranges of dependence, one efficient approach is wavelet transformation. It has the desirable property of representing an autocorrelated process with coefficients that are approximately uncorrelated \citep{flandrin92, fadili02}. In addition, there is no loss of information as wavelet bases are orthonormal, and the transformation is calculated with $\mathcal{O}(n \log_2 n)$ operations -- comparable to fast Fourier transform \citep{mallat09}. As detailed 
below, the Hurst parameter can be estimated from the decaying rate of the wavelet coefficients' variances across scales. 

Let $(\epsilon_1, \ldots, \epsilon_n)$ be a process with covariance function given by Equation~\ref{fgn}. The discrete wavelet transform decomposes the 
process into a set of 
detail coefficients $\{d_{jk}: j = 1, \dots, J, \ k = 1, \ldots, n/2^j\}$, and approximation coefficients $\{a_{Jk}: k = 1, \ldots, n/2^J\}$, where $J$ is the coarsest level of the transformation. Let $\gamma = 2H-1$, and $c_\gamma = (2\pi)^{-2H} \sin{(\pi H)} \Gamma(2H + 1)$. According to \cite{flandrin92, fadili02}, the covariance matrix of the wavelet transformed process is approximately diagonal, and the detail coefficients $\{d_{jk}: k = 1, \ldots, n/2^j\}$ and the approximation coefficients $\{a_{Jk}: k = 1, \ldots, n/2^J\}$ have variances
\begin{align}
\label{wavelet_variances}
S_{d_j} \approx \frac{\sigma^2 c_\gamma 2^{j\gamma} (2-2^\gamma)}{(2\pi)^\gamma (1-\gamma)} \; , 
\;
S_{a_J} \approx \frac{\sigma^2 c_\gamma 2^{(J+1)\gamma}}{(2\pi)^\gamma (1-\gamma)} \; .
\end{align}
Thus, $\gamma$, or equivalently the Hurst
parameter $H$, can be estimated 
through the linear relationship $\log_2(S_{d_j}) = \gamma j + \mathcal{O}(1)$. 
\begin{figure}[!tbh]
        \centering
        \includegraphics[width=\textwidth, height = .5\textwidth]{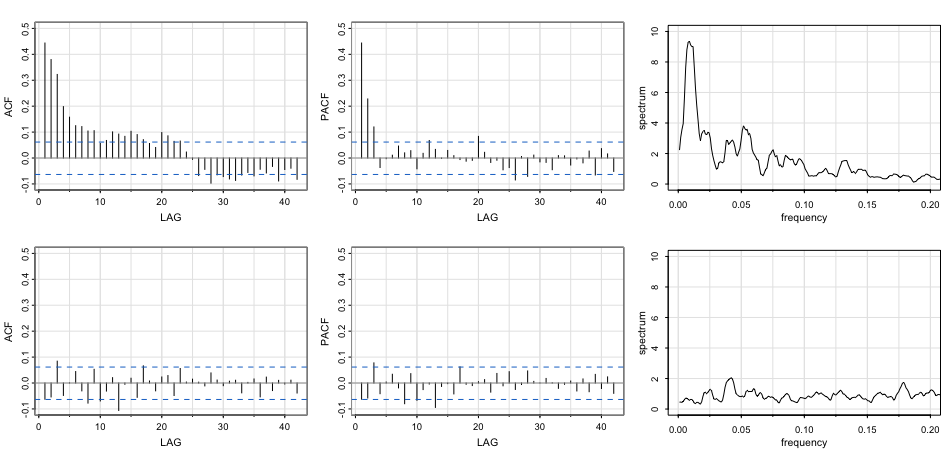}
        \caption{Autocorrelation function (left), partial autocorrelation function (middle) and spectrum-frequency plot (right) for a time
        series with long-range dependence (top) and its discrete wavelet transform (bottom).}
        \label{fig:dwt_decorrelation}
    \end{figure}
Figure~\ref{fig:dwt_decorrelation} demonstrates the decorrelating effect of the discrete wavelet transform on a time series with long range dependence. Compared to the original time series (top row), its wavelet transformation (bottom row) has insignificant autocorrelation coefficients, and much more evenly distributed spectral density across frequencies.

\subsection{The link between GMRFs and GFs}
\label{spde-lit}
In most analyses of interest, estimation 
requires calculations with
precision 
matrices ($\boldsymbol{Q} = \boldsymbol{\Sigma}^{-1}$),
not 
covariance 
matrices ($\boldsymbol{\Sigma}$). For example, if $\boldsymbol{\beta} = (\beta_1, \cdots, \beta_V) \sim N(\boldsymbol{\mu}, \boldsymbol{Q}^{-1})$, then the likelihood is
$$
L(\theta|\boldsymbol{\beta}) = \sqrt{(2\pi)^{-V} |\boldsymbol{Q}|} \; \exp{\bigg(-\frac{(\boldsymbol{\beta}-\boldsymbol{\mu})^\top \boldsymbol{Q}(\boldsymbol{\beta}-\boldsymbol{\mu})}{2}\bigg)}.
$$
Thus, GMRFs' sparse, banded precision matrix structure can significantly speed up model estimation.
By definition of GMRFs, such structure is a result of the conditional independence assumption.
Note that while it is entirely unrealistic to assume any two data points
to be independent, it is more reasonable to assume any two data points outside of their respective neighborhoods to be independent
{\em conditional on all others}. In this case, a sparse, banded precision matrix still has a dense inverse, i.e.~a dense covariance matrix. Even though estimation can be
performed speedily, there remains a serious challenge. Most GMRFs' precision matrices do not have an established correspondence with a closed-form covariance function. Real life data, including fMRI, are observed discretely from
continuous processes; and extrapolating beyond the discrete grid of 
observations requires knowing the functional form of the covariance. For example, for an unobserved vector $\boldsymbol{\beta}^*$, the conditional expectation 
$E(\boldsymbol{\beta}^*|\boldsymbol{\beta}) = \boldsymbol{\mu}_{\beta^*} - \boldsymbol{\Sigma}_{\beta^*\beta}\boldsymbol{Q}_{\beta\beta}(\boldsymbol{\beta}-\boldsymbol{\mu}_\beta)
$ is not available without knowing how to relate $\boldsymbol{Q}_{\beta\beta}$ to $\boldsymbol{\Sigma}_{\beta^*\beta}$. 

\citet{lindgren11} solved this problem for a class of covariance functions, namely the stationary Matérn
covariances. Their solution relies on a result by \citet{whittle54,whittle63} that links 
such class to the solutions of a particular stochastic partial differential equation (SPDE). Specifically, the stationary solutions $\beta(\boldsymbol{s})$ to the linear fractional SPDE
\begin{align}
\label{eq:spde}
(\kappa^2 - \Delta)^{\frac{\nu + d/2}{2}}\beta(\boldsymbol{s}) = \mathcal{W}(\boldsymbol{s}), \quad \boldsymbol{s} \in \mathbb{R}^d    
\end{align}
have Matérn covariance functions 
$$
C(\beta(\boldsymbol{0}), \beta(\boldsymbol{s})) = \frac{\sigma^2}{2^{\nu - 1} \Gamma(\nu)} (\kappa \norm{s})^\nu K_\nu(\kappa\norm{s})
$$
where $K_\nu(\cdot)$ is the modified Bessel function of the second kind and order $\nu$, $\Delta = \sum_{i = 1}^d \partial^2/\partial s_i^2$ is the Laplacian operator, $\mathcal{W}$ is Gaussian noise with unit variance, $\kappa$ is the spatial scale parameter, $\alpha = \nu + d/2$ controls the smoothness, and 
\begin{align*}
\sigma^2 = \frac{\Gamma(\nu)}{\Gamma(\nu + d/2)(4\pi)^{d/2}\kappa^{2\nu}}.  
\end{align*}
To complete their
proposal, \citet{lindgren11} formulated GMRF representations of the solutions to the SPDE in Equation~\ref{eq:spde}. The precision matrices 
$\boldsymbol{Q}$ of
such GMRF representations are explicitly parameterized by the SPDE parameters, or equivalently by the stationary Matérn covariance parameters. We start by giving a finite dimensional representation of the solutions to the SPDE; in symbols
\begin{align}
\label{eq:basis-expansion}
\beta(\boldsymbol{s}) = \sum_{l = 1}^L \psi_l(\boldsymbol{s})w_l    
\end{align}
for some deterministic basis functions $\{\psi_l\}$ and weights $w_l$. In our applications, the bases are piecewise linear, constructed 
by partitioning 
the domain into a set of non-intersecting triangles (i.e.~a triangular mesh). Specifically, given a triangle with vertices $i, j$ and $k$ located at $\boldsymbol{s}_i, \boldsymbol{s}_j$ and $\boldsymbol{s}_k$ in $\mathbb{R}^2$, for any inner location $\boldsymbol{s}$, $\psi_l(\boldsymbol{s})$ is equal to the area of the triangle formed by $\boldsymbol{s}$, $\boldsymbol{s}_j, \boldsymbol{s}_k$ divided by the area of the original triangle. It follows that $\psi_i$ takes a value of $1$ at the $i^{th}$ vertex and a value of $0$ at all other vertices. Let $\boldsymbol{C}, \boldsymbol{G}$, $\boldsymbol{K}$ and $\boldsymbol{Q_{\alpha}}$ be $L \times L$ matrices such that $C_{ij} = \langle \psi_i\psi_j \rangle$, $G_{ij} = \langle \nabla\psi_i\nabla\psi_j \rangle$, $K_{ij} = \kappa^2 C_{ij} + G_{ij}$ and
\begin{align}
\label{eq:Q}
\boldsymbol{Q}_\alpha = \begin{cases}
\boldsymbol{K}, \; \text{if } \alpha = 1\\
\boldsymbol{KC}^{-1}\boldsymbol{K}, \; \text{if } \alpha = 2\\
\boldsymbol{KC}^{-1}\boldsymbol{Q}_{\alpha-2}\boldsymbol{C}^{-1}\boldsymbol{K}, \; \text{if } \alpha = 3, 4, ...
\end{cases}
\end{align}
Then, if $\boldsymbol{w} = \{w_l\} \sim N(0, \boldsymbol{Q}^{-1}_{\alpha})$, the finite dimensional representations of the solutions to the SPDE in Equation \ref{eq:spde} are GFs with precisions $\boldsymbol{Q}_{\alpha}$. If $\boldsymbol{C}$ is replaced by the diagonal matrix $\tilde{\boldsymbol{C}}$ where $\tilde{\boldsymbol{C}}_{ii} = \langle \psi_i,1 \rangle$, we obtain GMRF representations instead of GFs. Note that $\nu$ (and hence $\alpha$) is usually fixed since it is typically not identifiable in applications. Thus, if $\boldsymbol{s} \in \mathbb{R}^2$, fixing $\nu = 1, 2,...$ gives $\alpha = 2, 3,...$ 

\subsection{Integrated Nested Laplace Approximation}
\label{inla}
For Bayesian hierarchical models that involve latent Gaussian fields, closed-form posterior distributions 
are, in general, unavailable. Sampling-based Markov-chain Monte Carlo (MCMC) methods are 
used instead of 
seeking analytical solutions.
While such methods are asymptotically exact,
in reality, 
computational cost and time may prevent one from achieving asymptotic guarantees; 
MCMC samples 
thus remain just repeated approximations of 
posterior distributions. The issue is particularly marked when MCMC methods are applied to latent Gaussian models; despite their 
versatility,
they can be 
extremely slow and the samples may not converge \citep{rue09}. To address this,
\citet{rue09} introduced the Integrated Nested Laplace Approximation (INLA), a method that performs direct numerical approximation of posterior distributions instead of sampling from them. 
Given observed data $\boldsymbol{y}$, the latent Gaussian field  $\boldsymbol{\beta} = \{\beta_v, v = 1,\ldots, V\}$ and a 
hyperparameter vector $\boldsymbol{\theta}$, the joint posterior distribution of $\boldsymbol{\beta}$ and $\boldsymbol{\theta}$ is 
\begin{align*}
\pi(\boldsymbol{\beta}, \boldsymbol{\theta}|\boldsymbol{y}) &\propto \pi(\boldsymbol{\theta})\pi(\boldsymbol{\beta}|\boldsymbol{\theta}) \pi(\boldsymbol{y}|\boldsymbol{\beta}, \boldsymbol{\theta})\\
&\propto \pi(\boldsymbol{\theta})|\boldsymbol{Q}(\boldsymbol{\theta})|^{1/2}\exp{\left[-\frac{1}{2}\boldsymbol{\beta}^\top \boldsymbol{Q}(\boldsymbol{\theta})\boldsymbol{\beta} + \log{(\pi(\boldsymbol{y}|\boldsymbol{\beta}, \boldsymbol{\theta}))} \right]}  \ .   
\end{align*}
Here, the goal is to approximate the posterior marginals 
\begin{align*}
\pi(\beta_v|\boldsymbol{y}) &= \int \pi(\beta_v,\boldsymbol{\theta}| \boldsymbol{y})d\boldsymbol{\theta} =
\int \pi(\beta_v|\boldsymbol{\theta}, \boldsymbol{y})\pi(\boldsymbol{\theta}|\boldsymbol{y})d\boldsymbol{\theta}
\propto \int \frac{\pi(\boldsymbol{\beta}, \boldsymbol{\theta}|\boldsymbol{y})}{\pi(\boldsymbol{\beta}_{-v}|\beta_v, \boldsymbol{\theta}, \boldsymbol{y})} \frac{\pi(\boldsymbol{\beta}, \boldsymbol{\theta}|\boldsymbol{y})}{\pi(\boldsymbol{\beta}| \boldsymbol{\theta}, \boldsymbol{y})}d\boldsymbol{\theta}\\
\pi(\theta_l|\boldsymbol{y}) &= \int \pi(\boldsymbol{\theta}|\boldsymbol{y}) d\boldsymbol{\theta}_{-l} \propto \int \frac{\pi(\boldsymbol{\beta}, \boldsymbol{\theta}|\boldsymbol{y})}{\pi(\boldsymbol{\beta}| \boldsymbol{\theta}, \boldsymbol{y})}d\boldsymbol{\theta}_{-l}
\end{align*}
where $\boldsymbol{\beta}_{-v} = \boldsymbol{\beta}\setminus \{\beta_v\}$ and $\boldsymbol{\theta}_{-l} = \boldsymbol{\theta}\setminus \{\theta_l\}$. The computational cost of numerical integration increases significantly with the dimension of the hyperparameter vector $\boldsymbol{\theta}$;
INLA can accommodate up to 
about $10$ hyperparameters, but becomes unfeasible in higher dimension. For a given $\boldsymbol{\theta}$, the denominator densities $\pi(\boldsymbol{\beta}_{-v}|\beta_v, \boldsymbol{\theta}, \boldsymbol{y})$ and $\pi(\boldsymbol{\beta}| \boldsymbol{\theta}, \boldsymbol{y})$ are approximated by $\tilde{\pi}(\boldsymbol{\beta}_{-v}|\beta_v, \boldsymbol{\theta}, \boldsymbol{y})$ and $\tilde{\pi}(\boldsymbol{\beta}| \boldsymbol{\theta}, \boldsymbol{y})$, respectively, using Laplace approximation. The marginals $\pi(\beta_v|\boldsymbol{y})$ and $\pi(\theta_l|\boldsymbol{y})$ are then approximated by 
numerical integration over $\boldsymbol{\theta}$; that is
 \begin{align*} 
 \pi(\beta_v|\boldsymbol{y}) &\approx \int \frac{\pi(\boldsymbol{\beta}, \boldsymbol{\theta}|\boldsymbol{y})}{\tilde{\pi}(\boldsymbol{\beta}_{-v}|\beta_v, \boldsymbol{\theta}, \boldsymbol{y})} \frac{\pi(\boldsymbol{\beta}, \boldsymbol{\theta}|\boldsymbol{y})}{\tilde{\pi}(\boldsymbol{\beta}| \boldsymbol{\theta}, \boldsymbol{y})}d\boldsymbol{\theta}\\
\pi(\theta_l|\boldsymbol{y}) &\approx \int \frac{\pi(\boldsymbol{\beta}, \boldsymbol{\theta}|\boldsymbol{y})}{\tilde{\pi}(\boldsymbol{\beta}| \boldsymbol{\theta}, \boldsymbol{y})}d\boldsymbol{\theta}_{-l} \ .
 \end{align*} 
 The computational 
 speed-up provided by INLA with respect to 
 MCMC algorithms 
 is on the scale of seconds/minutes compared to hours/days, 
 with comparable approximation errors. For more details, see \citet{rue09}.

\section{Modeling approach}
\label{sec:modeling_approach}

A standard model for an fMRI
data set is
\begin{align}
\label{eq:c4model}
y_{vt} = \sum_{k = 0}^K x_{kt}\beta_{kv} + \epsilon_{vt}    
\end{align}
where $v \in \{1,\cdots, V\}$ indexes location, 
i.e.~vertices in cs-fMRI, $t \in \{1, \cdots, T\}$ indexes time, and $k \in \{1, \cdots, K\}$ indexes a task or stimulus. 
$x_{kt}$ is the convolution of a so-called stimulus time course 
$s_k(\cdot)$ for task $k$ with the canonical hemodynamic response function 
$h(\cdot)$; namely
\begin{align}
\label{x_definition}
    x_{kt} = \int_{-\infty}^\infty h(u)s_k(t-u)du \ .
\end{align} 
$s_k(\cdot)$ takes value $1$ when the
task is active, 
and $0$ otherwise. 
The canonical hemodynamic response function $h(\cdot)$, visualized in Figure~\ref{fig:simulation-design}, characterizes the temporal change in oxygenated blood flow 
for regions of the brain that are affected by 
the task. 
This typically 
consists of a 2 seconds delay at the onset of the stimulus, then a dip below baseline followed by a gradual increase that peaks after 4 seconds, a slow decay to below baseline level, and a return to baseline level. The duration of each phase may vary 
depending on the 
task, and the total amount of response time is approximately 15 to 20 seconds \citep{lazar08}. 
We can rewrite the above model for all $V$ vertices and $T$ time points in vector form as
\begin{align}
\label{design}
\boldsymbol{y} = \sum_{k = 0}^K \boldsymbol{X}_k\boldsymbol{\beta}_k + \boldsymbol{\epsilon}, \quad \boldsymbol{\epsilon} \sim N(0, \boldsymbol{\Sigma}) 
\end{align}
%
where $\boldsymbol{y}$ is now a $VT \times 1$ vector
created by stacking $V$ time series, each of length $T$, 
$\boldsymbol{X}_k$ is a $VT \times V$ design matrix containing 
activation information for 
task 
$k$, and $\boldsymbol{\beta}_k$ is a $V \times 1$ vector of activation amplitudes. 
As we shall see, the spatial dependence is modeled via non-stationary Matern priors on the task activation fields $\boldsymbol{\beta}_k$. The $VT \times VT$ error covariance matrix $\boldsymbol{\Sigma}$ is block diagonal, with $V$ $T \times T$ blocks $\{\boldsymbol{\Sigma}_v: v = 1, \dots, V\}$, each capturing the temporal dependence at a particular vertex.

\subsection{Modeling varying-range temporal dependence}
Temporal dependence can be modeled separately at each vertex, as the covariance matrix $\boldsymbol{\Sigma}$ in Equation~\ref{design} is block diagonal. To this end,
the fMRI time series at a generic vertex $v$ is modeled as a fractional Gaussian noise process, whose covariance $\boldsymbol{\Sigma}_v$ is parameterized by the Hurst parameter $H_v$ (see 
Section~\ref{sec:dwt}). After
discrete wavelet 
transformation of the time series at vertex $v$, 
Equation~\ref{eq:c4model} becomes
$$
\boldsymbol{y}_{v}^{(w)} = \sum_{k = 0}^K \boldsymbol{x}^{(w)}_{k}\beta_{kv} + \boldsymbol{\epsilon}_v^{(w)}, \quad \boldsymbol{\epsilon}_v^{(w)} \sim N(0, \boldsymbol{\Sigma}_v^{(w)})
$$
where $\boldsymbol{x}_k^{(w)}$, $\boldsymbol{y}_v^{(w)}$ and $\boldsymbol{\epsilon}_v^{(w)}$ are the discrete wavelet transforms of 
$x_{kt}, y_{vt}$ and $\epsilon_{vt}$,
$t = 1, \ldots, T$.
Because of the approximately decorrelating property of the discrete wavelet transformation, the covariance matrix $\boldsymbol{\Sigma}_v^{(w)}$ is, to a good approximation, diagonal. As described in 
Section~\ref{sec:dwt}, the diagonal entries depend on the scale of wavelet transform, and on the range of temporal dependence at vertex $v$.  With the goal of accommodating different ranges of dependence across brain regions of interest, while keeping the model parsimonious, we devise a data-driven scheme that
groups the ranges
into just $n_H$ ($<< V$) distinct levels, thus leaving us with only $n_H$ Hurst parameters to estimate. We proceed as 
follows:
\begin{enumerate}[label = Step~3.1.\arabic*]
\item At each vertex $v$, we obtain 
preliminary estimates of coefficients $\{\hat{\beta}_{kv}:  k = 1, \ldots, K\}$ 
and residuals $\{\hat{\epsilon}_{vt}: t = 1, \ldots, T\}$  using linear regression;
\item At each vertex $v$, we take the discrete wavelet transform of the residuals $\{\hat{\epsilon}_{vt}: t = 1, \ldots, T\}$, producing
detail coefficients $\{d_{jk}: j = 1, \dots, J, k = 1, \ldots, n/2^j\}$
and approximation coefficients $\{a_{Jk}: k = 1, \ldots, n/2^J\}$, where $J$ is the coarsest scale of transformation;
\item At each vertex $v$, we estimate the variances $\{S_{d_j}: j = 1, \dots, J\}$ in 
Equation~\ref{wavelet_variances} 
computing the variances of $\{d_{jk}: j = 1, \ldots, n/2^j\}$;
\item At each vertex $v$, we estimate $\gamma$ via the linear relationship $\log_2(S_{d_j}) = \gamma j + \mathcal{O}(1)$
and produce a preliminary estimate of the Hurst parameter as
$\hat{H}_v = (\hat{\gamma} + 1)/2$;
\item For each region of interest $r = 1, \ldots, R$, we obtain the median of such 
such estimates across vertices in the region,
$M_r^{(H)} = \text{med} \{
\hat{H}_v: v \in r\}$;
\item 
We group 
the $R$ regions into $n_H$ clusters based on their median estimates \mbox{$\{M_r^{(H)}: r = 1, \ldots, R\}$},
producing cluster memberships $C_r \in \{1, \ldots, n_H\}$ for each 
region $r$. Since no two regions share a vertex, 
cluster memberships 
$C_v \in \{1, \ldots, n_H\}$ can also be attributed to each individual vertex
$v \in \{1, \ldots, V\}$ based on the membership of the regions it belongs to;
\item At each vertex $v$, we assume the 
time series 
to be distributed as a fractional Gaussian process
parameterized by the cluster-specific Hurst parameter corresponding to the vertex membership.
\end{enumerate}
At the end of this procedure, any two
vertices $v$ and  $v'$ such that $C_v = C_{v'}$
will have the same Hurst parameter, and thus their $\boldsymbol{\Sigma}_v^{(w)}$ and $\boldsymbol{\Sigma}_{v'}^{(w)}$ will have the same diagonal entries expressed in 
Equation~\ref{wavelet_variances}. Note that we are not using the  procedure to estimate the parameters, we are just employing rough, preliminary estimates $\hat{H}_v$ to group regions and vertices, and thus reduce the number of parameters to be estimated.
Note also that, 
in Step~3.1.2, the number of available coefficients to estimate $S_{d_j}$ is $n/2^j$. As $j$ increases, this number decreases by a factor of 2, causing estimation of $S_{d_j}$ to be volatile at large scales. Hence, we recommend using only 
scales with at least 16 
detail coefficients to generate the preliminary 
$\hat{H}_v$'s in Step~3.1.3.
Finally, we implement the 
clustering in Step~3.1.6 
using a simple 
K-mean algorithm with the
number of clusters $n_H$ 
fixed beforehand
(more clusters correspond to more distinct 
dependence ranges, and thus to a less parsimonious model, with more parameters to be estimated). While more sophisticated approaches could be employed, 
in our experience a K-means 
with 
$n_H \leq 5$ works sufficiently well (larger $n_H$ values 
do not provide any added improvements in model fit). The choice to perform clustering on regions instead of directly on vertices is based on the assumption that vertices in the same region have similar temporal dependence. This 
is supported by \citet{park10}, in which the authors 
estimated 
Hurst parameters 
separately for each brain voxel, after removing spatial covariance from the data;
even with spatial covariance removed, 
Hurst parameter estimates
still mapped nicely into
obvious brain structures.
Clustering 
regions has the added advantage of reducing spurious temporal noise, which may arise naturally in the data, and/or from previous estimation steps in the algorithm. 
We also note that, while beyond the scope of the current 
article, 
size and interpretation of the regions being clustered can be flexibly altered by changing the choice of parcellation.

\begin{figure}[!tbhp]
        \centering
        \includegraphics[width=\textwidth]{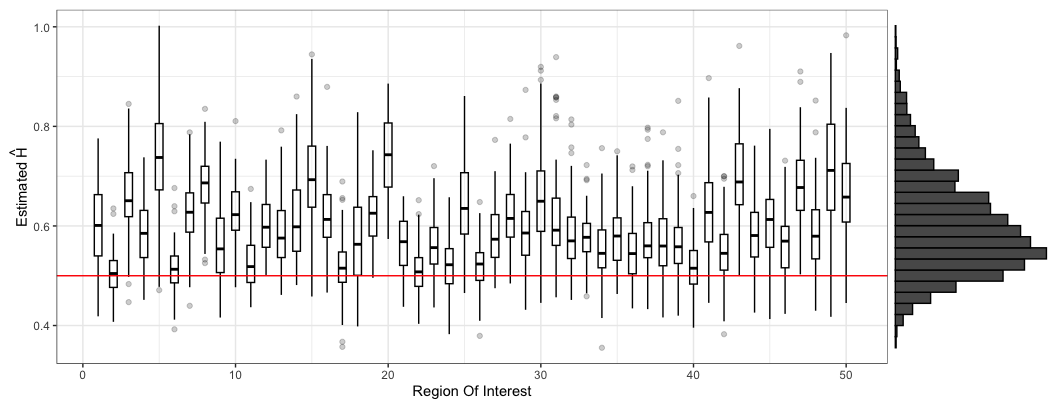} 
        \caption{Boxplots of
        $\hat{H}_v$
        values grouped by $50$ left hemisphere cortical regions from the Schaefer parcellation \citep{schaefer18}. The numbering of the regions does not correspond to any particular arrangement or order.
        The histogram on the secondary y-axis (right) 
        shows the overall distribution. The horizontal red line
        marks $H = 0.5$, which 
        corresponds to a white noise process. 
        }
        \label{fig:initial_H_by_roi}
\end{figure}
Figure~\ref{fig:initial_H_by_roi} provides an illustration of the 
preliminary estimates $\hat{H}_v$ 
which are represented as boxplots for each 
region of interests
in the Schaefer 100 parcellation 
\citep[see Figure~\ref{fig:surface},
each hemisphere's cortical surface is partitioned into 50 regions;][]{schaefer18}.
Specifically, the illustration uses resting state time series ($1,024$ time points) 
for $6,000$ vertices in the left hemisphere of an individual from the healthy young adult data set of the Human Connectome Project 
\citep[HCP;][]{vanessen13}.
Of course, in task-related (as opposed to resting state) fMRI data, 
temporal dependence, and thus Hurst parameter estimates,
may be influenced by the task.
%
The horizontal red line in Figure~\ref{fig:initial_H_by_roi} 
marks $H = 0.5$, which corresponds to a white noise process. 
Although many 
regions have similar
$\hat{H}_v$ values (e.g.,~regions 32--39), we
do see evidence of varying ranges of dependence (values between $H = 0.5$ and $H = 1$) even in this resting state data.
The fact that for some regions almost all coarse estimates
are above $H=0.5$ (e.g.,~regions 3, 8, 12, 20, 43 and 47), together with the skew of the overall distribution (histogram on the right of Figure~\ref{fig:initial_H_by_roi}), strongly suggest the existence of
long-range dependence. 

\subsection{Modeling non-stationary spatial dependence}
\label{spatial prior}
In this section, the task indices are suppressed for convenience, so that $\boldsymbol{\beta}_k = \boldsymbol{\beta}$ and $\beta_{kv} = \beta_{v}$. The SPDE framework can be extended to accommodate non-stationary processes by introducing location-dependent 
SPDE parameters.
Following \citet{lindgren11}, a parameter $\tau$ is used to scale $\beta$ in the SPDE
in~(\ref{eq:spde}), while keeping the variance of the Gaussian noise constant; that is
\begin{align*}
(\kappa^2(\boldsymbol{s}) - \Delta)^{\frac{\nu + d/2}{2}}\tau(\boldsymbol{s})\beta(\boldsymbol{s}) = \mathcal{W}(\boldsymbol{s}), \quad \boldsymbol{s} \in \mathbb{R}^d    
\end{align*}
where $\boldsymbol{s}$ indicates the location in a space of generic dimension $d$. Note the distinction between this continuous coordinates notation, which is necessary for distance calculations, and the vertex indices $v \in \{1, \ldots, V\}$.
The marginal variance becomes
\begin{align}
\label{eq:matern_variance}
\sigma^2(\boldsymbol{s}) = \frac{\Gamma(\nu)}{\Gamma(\nu + d/2)(4\pi)^{d/2}\kappa^{2\nu}(\boldsymbol{s})\tau(\boldsymbol{s})} \ . 
\end{align}
As noted in \citet{lindgren15}, 
it is often more intuitive to 
reparameterize the SPDE
using the standard deviation $\sigma$ and the range $\rho = (8\nu)^{0.5}/\kappa$, where $\rho$ is the distance at which 2 observations are approximately independent (correlation 
approximately $0.13$) for all $\nu > 0.5$. We model $\sigma(\boldsymbol{s})$ and
$\rho(\boldsymbol{s})$ as functions of 
the location and translate them back to $\tau(\boldsymbol{s})$ and $\kappa(\boldsymbol{s})$. Specifically, we set
\begin{align}
\begin{split}
\label{model_sigma_rho}
    &\log(\sigma(\boldsymbol{s})) = \log(\sigma_0) + \theta_1 \delta(\boldsymbol{s})\\
    &\log(\rho(\boldsymbol{s})) = \log\rho = \log(\rho_0) + \theta_2 
\end{split}
\end{align}
where $\sigma_0$ and $\rho_0$ are baseline standard deviation and range, and $\delta(\boldsymbol{s})$ is some chosen local variability score. 
Using the relationship $\rho = (8\nu)^{0.5}/\kappa$ and Equation~\ref{eq:matern_variance} we thus get
\begin{align*}
    \log \kappa(\boldsymbol{s}) 
    &= \log \kappa = \frac{\log (8\nu)}{2} - \log(\rho_0) - \theta_2 \\
    &= \log\kappa_0 - \theta_2 \\
    \log \tau(\boldsymbol{s}) 
    &= \frac{1}{2}\log\left(\frac{\Gamma(\nu)}{\Gamma(\nu + \frac{d}{2})(4\pi)^{\frac{d}{2}}}\right) - \log(\sigma_0) - \nu \left(\frac{\log (8\nu)}{2} - \log(\rho_0)\right) 
    - \theta_1\log(\delta(\boldsymbol{s})) + \theta_2\nu \\ 
    &= \log\tau_0 - \theta_1\delta(\boldsymbol{s}) + \theta_2\nu \ .
\end{align*}
Let $\boldsymbol{T} = \text{diag}(\tau(\boldsymbol{s_v}): v = 1, \ldots, V)$, then the precision matrix in 
Equation~\ref{eq:Q} becomes $\boldsymbol{T}\boldsymbol{Q}_\alpha\boldsymbol{T}$. \cite{lindgren11, lindgren15} remarked that the link between SPDE and Matern parameters is no longer valid in the non-stationary adaptation. However, when $\kappa(\boldsymbol{s})$ and $\tau(\boldsymbol{s})$ vary slowly over the domain, 
Equations~\ref{model_sigma_rho} provide a valid approximation to the local variances and correlation ranges. This requirement is achieved if the local variability score $\delta(\boldsymbol{s})$ is also slowly varying with $\boldsymbol{s}$. For our applications, we 
calculate the local variability score as 
follows:
\begin{enumerate}[label = Step~3.2.\arabic*]
\item 
At each vertex $v$, we obtain initial estimates $\hat{\beta}_v$ using linear regression;
\item 
At each vertex $v$,
we calculate 
$\delta_v = \delta(\boldsymbol{s}_v)$ 
as the standard deviation of the values $\hat{\beta}_i$, $i \in N_v$,
where $N_v$ is a neighborhood of $v$. 
\end{enumerate}
The size of the neighborhood $N_v$ in Step~3.2.2 can be chosen such that $\delta(\boldsymbol{s})$ is smooth in $\boldsymbol{s}$. For sufficiently dense data such as ours, nearest neighbors suffice.

\subsection{Tying it all together: joint posterior}
Given the wavelet transformed data $\boldsymbol{y}^{(w)}$, we can now write the joint posterior of 
marginal temporal variance 
$\sigma$, 
temporal Hurst parameters $H_1, \ldots, H_{n_H}$ (recall these are reduced in number with respect to the number of vertices
and are not task dependent)
and, for each task $k = 1, \dots, K$, 
activation fields $\boldsymbol{\beta}_k = \{\beta_{k1},\ldots, \beta_{kV}\}$
and 
spatial hyperparameters $\boldsymbol{\theta}_k = \{\theta_{k1}, \theta_{k2}\}$. Such joint posterior is
\begin{align*}
    &\pi(\boldsymbol{\beta}_1,\ldots, \boldsymbol{\beta}_K, \boldsymbol{\theta}_1, \ldots, \boldsymbol{\theta}_K, H_1,\ldots, H_{n_H}, \sigma|\boldsymbol{y}^{(w)}) \\
    &\propto \pi(\sigma) \left[\prod_{k = 1}^{K}\pi(\boldsymbol{\theta}_k)\pi(\boldsymbol{\beta}_k|\boldsymbol{\theta}_k)\right] \left[\prod_{i = 1}^{n_H}\pi(H_i)\right] \left[\prod_{v = 1}^V \pi(\boldsymbol{y}_v^{(w)}|\boldsymbol{\beta}_1,\ldots, \boldsymbol{\beta}_K, H_1,\ldots, H_{n_H}, \sigma)\right]\\
    &\propto \pi(\sigma) \left[\prod_{k = 1}^{K}\pi(\boldsymbol{\theta}_k)|\boldsymbol{Q}(\boldsymbol{\theta}_k)|^{1/2}\exp\left(-\frac{1}{2}\boldsymbol{\beta}_k^\top \boldsymbol{Q}(\boldsymbol{\theta}_k)\boldsymbol{\beta}_k\right)\right] \left[\prod_{i = 1}^{n_H}\pi(H_i)\right] \times \\
    &\hspace{1.2cm} \left[\prod_{v = 1}^V |\boldsymbol{\Sigma}_v^{(w)}|^{-1/2} \exp\left(\frac{1}{2} \bigg(\boldsymbol{y}_v^{(w)} - \sum_{k = 0}^K \boldsymbol{x}^{(w)}_{k}\beta_{kv}\bigg)^\top {\boldsymbol{\Sigma}_v^{(w)}}^{-1}\bigg(\boldsymbol{y}_v^{(w)} - \sum_{k = 0}^K \boldsymbol{x}^{(w)}_{k}\beta_{kv}\bigg)\right)\right] 
\end{align*}
where $\pi(\sigma)$ is Gamma(1,1); $\{\pi(H_i): i = 1, \ldots, n_H\}$ are Uniform(0,1); $\{\pi(\theta_{kj}): k = 1, \ldots, K;\  j = 1, 2\}$ are Gaussian with mean $0$ and precision $0.3$;  $\{\pi(\boldsymbol{\beta}_k|\boldsymbol{\theta}_k):  k = 1, \ldots, K\}$ are 
GMRFs with mean $0$ and precision $\boldsymbol{Q}(\boldsymbol{\theta}) = \boldsymbol{T}\boldsymbol{Q}_\alpha\boldsymbol{T}$, as described in 
Section~\ref{spatial prior}.
Based on this joint posterior distribution, marginal posteriors for different parameters are approximated using INLA, as described in
Section~\ref{inla}. We shall call our approach \emph{non-stationary, varying-range Bayesian General Linear Model}, hereinafter \emph{NSVR-BayesGLM}. 

\section{Results\label{sec:3}}
\subsection{Simulations}
\label{c4simulations}
\begin{figure}[!hbtp]
    \centering
    \includegraphics[width=\textwidth]{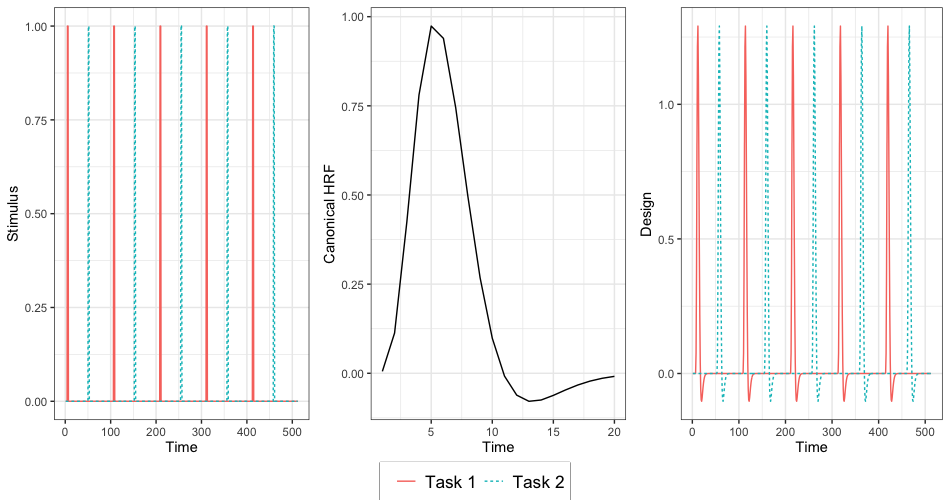} 
    \caption{Stimulus function (left), canonical hemodynamic response function (middle), and the design matrix that results from their convolution (right).}
    \label{fig:simulation-design} 
\end{figure}

 Since the main target of our proposed approach are cs-fMRI data, which provide a rendition of the cerebral cortex via a 2D surface mesh, it makes sense to consider 2D simulations. Here, to facilitate comparisons with prior literature, we 
 build simulations on a traditional 2D brain slice. Specifically, 
 we follow the set-up of \citet{mejia20}
 and compare results between our \emph{NSVR-BayesGLM} and their \emph{BayesGLM} approach.
A $46 \times 55$ image is constructed from a brain mask provided by the \textsf{SPM8} software. Activation is placed at 
four different sites,
with varying signal strengths, spreads and degrees of smoothness.
Signal strength is defined as the ratio of activation magnitude 
at the center of 
the site, $M = |\beta_{center}|$,  
to 
noise standard deviation, $\sigma$.
Spread is defined as the radius $r$ of the 
site. 
Smoothness is defined by how fast the signal decays from the center to the edge of 
the site, via the exponential function 
$e^{-\lambda d}$,
where $d$ is the distance to the center and $\lambda$ the parameter that controls smoothness. Thus, for an activation site with signal strength $M/\sigma$, radius $r$ and smoothness parameter $\lambda$, the activation magnitude at a location $d$ distance away from the center is 
$Me^{-\lambda d}$
if $d < r$ and 0 otherwise. 
Then, at different sites, we add 
error time series 
with different dependence ranges. Specifically, we place fractional Gaussian noise processes with $H = 0.8$ in two active regions to represent long temporal dependence, $H = 0.4$ in the remain two active regions to represent short temporal dependence, and $H = 0.5$ outside of active regions to represent white noise. We assume no overlapping between activation sites. For convenience, we choose $\sigma = 1$ and, 
since the discrete wavelet transform requires $T$ to be a power of two, $T = 2^9 = 512$. In reality, for 
any $T$, one can perform some ``padding'' 
to meet this requirement.
A summary of the simulation design can be found in 
Table~\ref{table:sim-design} --
the two tasks, which are depicted in Figure~\ref{fig:simulation-design}, alternate in a single-block design scheme.
\begin{table}[!b]
\centering
\begin{tabular}{|c|c|c|c|c|} 
\hline
 & Signal strength & Activation center  & Smoothness & H \\
\hline
&& top &0.2&0.8\\
\cline{3-5}
&& right &0.2&0.4\\
\cline{3-5}
&& bottom &0.05&0.4\\
\cline{3-5}
\multirow{-4}*{Task 1}& \multirow{-4}*{2} &  left &0.05&0.8\\
\hline
&& top &0.2&0.8\\
\cline{3-5}
&&right &0.2&0.4\\
\cline{3-5}
&&bottom &0.05&0.4\\
\cline{3-5}
\multirow{-4}*{Task 2}&\multirow{-4}*{3}& left &0.05&0.8\\
\hline
\end{tabular}
 \caption{Details of simulation design.}
 \label{table:sim-design}
\end{table}

Both 
\emph{NSVR-BayesGLM} and \emph{BayesGLM} use SPDE priors to model spatial dependence of the activation fields,
and both are implemented with the \textsf{R} package \textsf{R-INLA}
using the parallel \textsf{PARDISO} solver \citep{rue13}. 
The varying temporal dependence component of 
\emph{NSVR-BayesGLM} is written and implemented in C as a custom add-on that operates within \textsf{R-INLA}. While \emph{BayesGLM} prewhitens all time-series with an $AR(6)$ model and assumes stationary spatial priors, \emph{NSVR-BayesGLM} 
allows varying ranges of temporal dependence in wavelet space and 
non-stationary spatial priors in one, unifying framework. 

\begin{figure}[!t]
    \centering
    \includegraphics[width=\textwidth]{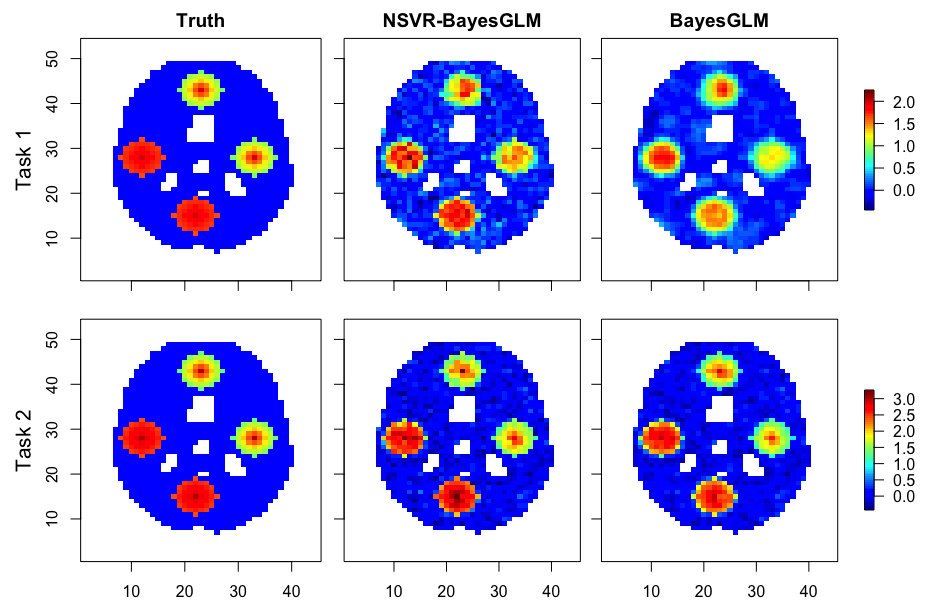} 
    \caption{Simulated data (left) and activation magnitude estimates by NSVR-BayesGLM (middle) and BayesGLM (right). 
    Top and bottom plots 
    concern Tasks 1 and 2, respectively.
    The ventricles, which house cerebral spinal fluid and are located near the center of the brain, are plotted in white and are not analyzed.}
    \label{fig:sim-result} 
\end{figure}

Figure~\ref{fig:sim-result}
shows the 
true activation magnitudes 
(left) and
estimates by \emph{NSVR-BayesGLM} (middle) and \emph{BayesGLM} (right), for all simulation settings. Different gradients of colors at activation sites portray different degrees of spatial smoothness. Whereas the non-stationary feature of \emph{NSVR-BayesGLM} allows it to capture this varying spatial smoothness, \emph{BayesGLM}'s stationary priors assume the same smoothness level across all locations. This is evident in the latter's tendency to oversmooth activation magnitudes in the bottom and the left activation 
sites. \emph{BayesGLM} also underestimates activation magnitudes, which is most evident at the centers of activation 
sites.
This is likely a combined effect of prewhitening with $AR(6)$ models, as 
discussed in Section~\ref{sec:temporal_dependence}, and of assuming stationarity. 
In addition to the activation magnitudes captured in Figure~\ref{fig:sim-result}, each activation
site has a different temporal range of dependence. 
Here we use 
$n_H = 3$;
\emph{NSVR-BayesGLM} automatically groups regions 
with similar temporal dependence ranges -- such as the right and bottom activation 
sites -- and produces Hurst parameter estimates $\hat{H} = 0.373, 0.825, 0.486$ for the true $H = 0.4, 0.8, 0.5$, respectively. 
Note that, if instead we use $n_H = 5$, \emph{NSVR-BayesGLM} 
produces estimates $\hat{H} = 0.390, 0.396, 0.853, 0.871, 0.487$ for the true $H = 0.4, 0.4, 0.8, 0.8, 0.5$, respectively. 
\emph{BayesGLM} does not accommodate varying ranges of dependence and therefore does not 
produce estimates of such ranges. 

\begin{figure}[!bt]
    \centering
    \includegraphics[width=\textwidth]{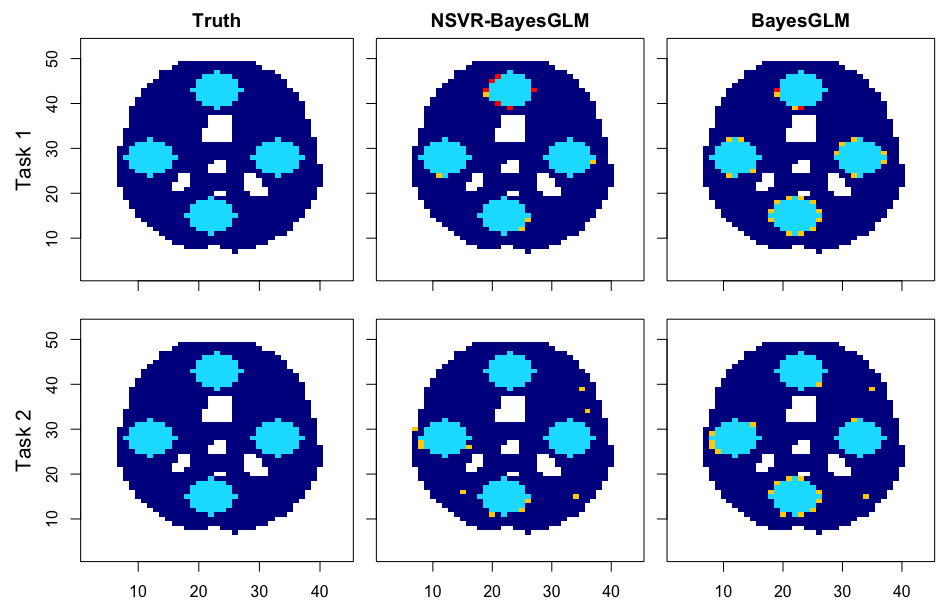} 
    \caption{Simulated data (left) and activation region estimates by \emph{NSVR-BayesGLM} 
    (middle) and \emph{BayesGLM} (right). 
    Top and bottom plots 
    concern 
    Tasks 1 and 2, respectively. 
    False positives are plotted in yellow
    and false negatives 
    in red
    (as in Figure~6, ventricles
    are plotted in white and are not analyzed).
    }
    \label{fig:sim-result-active-regions} 
\end{figure}
Figure~\ref{fig:sim-result-active-regions} shows
the true activation 
sites (left), 
and 
estimates by \emph{NSVR-BayesGLM} (middle) and \emph{BayesGLM} (right), for all simulation settings. In both 
approaches, an estimated activation 
site is determined as the largest set $D$ of 
contiguous locations such that the joint posterior probability $P(\beta_v > 0
\text { for all } v \in D \,|\,\boldsymbol{y}) > 1 - \alpha$, where $\alpha$ is
a posterior probability threshold
which we fixed at $0.05$.
This way of estimating activation 
sites is called the {\em excursions set approach};
it was introduced by \cite{bolin15}
and is implemented in the \textsf{R} package \textsf{excursions} \citep{bolin18}. As detailed in \cite{mejia20}, since the joint posterior probability is over all vertices, 
there is no need to adjust for multiple comparisons -- 
and the approach appears to be fairly insensitive to the choice of $\alpha$. 
Based on the top 
plots of Figure~\ref{fig:sim-result-active-regions}, in 
Task~1, both \emph{NSVR-BayesGLM} and \emph{BayesGLM} 
produce false negatives around the edge of the top and right activation
sites, likely an effect of low signal strength 
in those areas. Notably, \emph{BayesGLM} is also particularly prone to false positives, even in the case of strong signals (Task 2, bottom plots).
The 
stationarity assumption of \emph{BayesGLM} acts like a smoother of activation magnitudes, dampening those that are larger and boosting those that are smaller. This has the effect of making false positives more likely around the edges of the bottom and left activation
sites, while making false negatives less likely around the edges of the top and right 
activation sites.

\subsection{An application to visual working memory}
\label{sec:application}
In this application, we use cs-fMRI data to
characterize 
activation and temporal dependence 
in a visual working memory task.
The data is part of a working memory study included in the healthy young adult data set of the
Human Connectome Project 
\citep[HCP;][]{vanessen13}. 
Preprocessing is conducted with the HCP
minimal preprocessing pipeline \citep{glasser13}, and individual subjects' cortical areas are registered to a common template for the purpose of group analysis and intersubject comparisons \citep{robinson14}. In the
study, subjects are presented with blocks of trials 
consisting of different images. Subjects indicate if an image is a 2-back repeat (i.e.,~occurred
2 trials before) in a `2-back' condition, or if an image matches 
a cue stimulus in a `0-back' condition (as a control for working memory). In summary, 
four `2-back' blocks and 
four `0-back' blocks alternate randomly, for a total of 
eight blocks per run,
each block 
consisting of 10 trials of 2.5 seconds each. For all subjects, a working memory score is also measured, using the NIH Toolbox List Sorting Working Memory Test \citep{tulsky14}. For the purposes of our analysis,
we select 10 subjects from the 
100 unrelated subjects (no family relations) available from the HCP; namely, the 5 highest- and the 5 lowest-ranking 
in the List Sorting Working Memory Test.

We analyze 
left and right hemispheres separately, using the approach described in 
Section~\ref{sec:modeling_approach}. To reduce computational burden, we use the Connectome Workbench 
to subsample
vertices in each hemisphere from $32,492$ 
to 
$\approx 3,000$. Thus, for our modeling exercise
$V \approx 3,000$ and $T = 401$. The BOLD response and design matrix are centered and scaled at each vertex for numerical stability. In addition, 
spurious variability caused by subject movements and scanner drift is
removed from the BOLD signal, using $6$ motion parameters estimates (translation and rotation parameters, each in 3 directions) and their temporal derivatives as regressors. We carry out the modelling
on the `very-inflated' surface, and 
visualize results on the `inflated' surface (see Figure~\ref{fig:surface}). As in 
Section~\ref{c4simulations},
\emph{NSVR-BayesGLM} is implemented with the \textsf{R} package \textsf{R-INLA}, using the parallel \textsf{PARDISO} solver \citep{rue13}, and with the varying temporal dependence component 
written in C as a custom add-on 
within \textsf{R-INLA}. Manipulation of cortical surface fMRI data in \textsf{R} is made possible 
through the \textsf{ciftiTools} package \citep{pham22}. Using 8 parallel threads, computation takes on average 30 minutes per hemisphere per subject, on a Macbook Pro with M2 Max chip and 96GB memory.
 \begin{figure}[!tbhp]
        \centering
        \includegraphics[width=\textwidth]{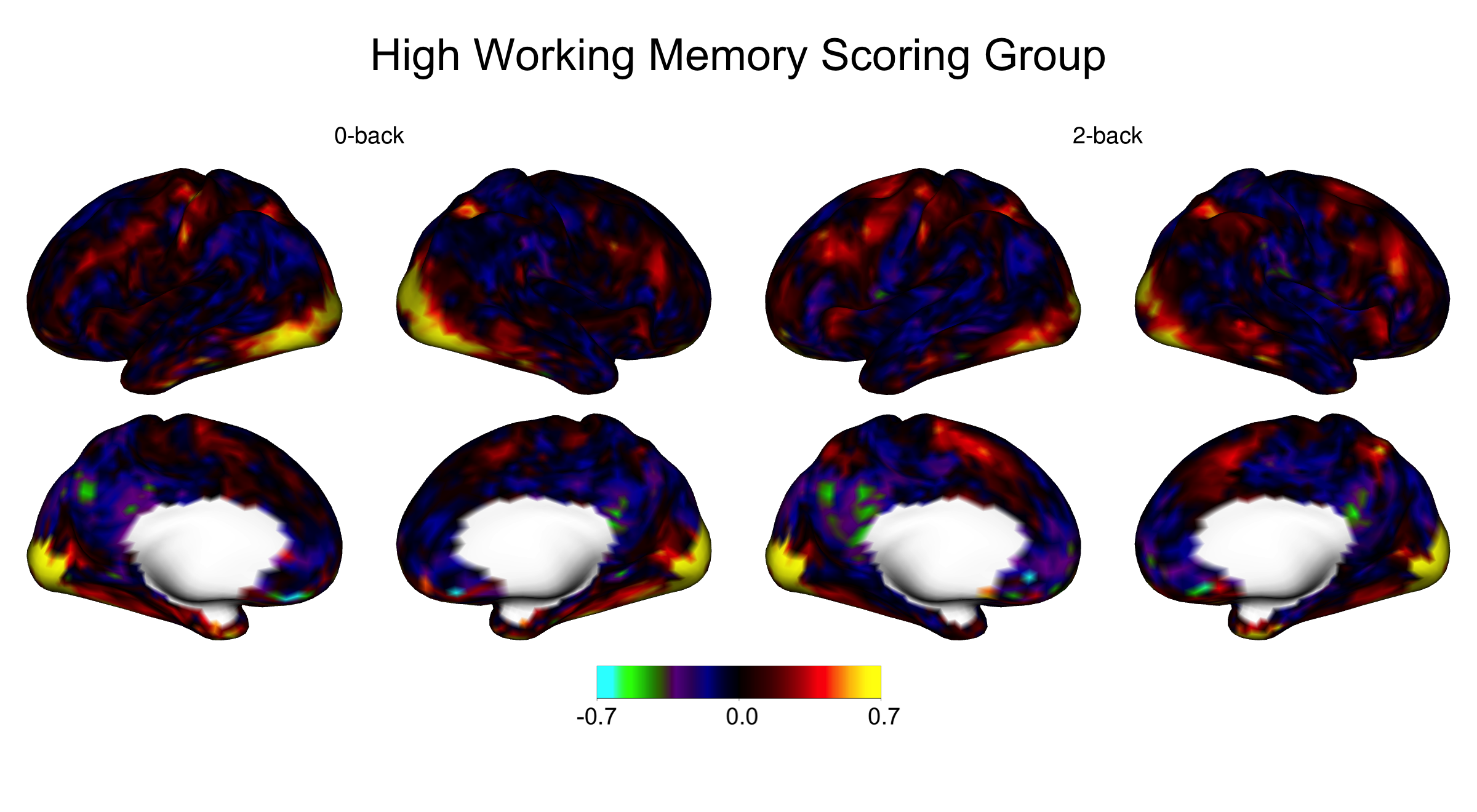}\\
        \includegraphics[width=\textwidth]{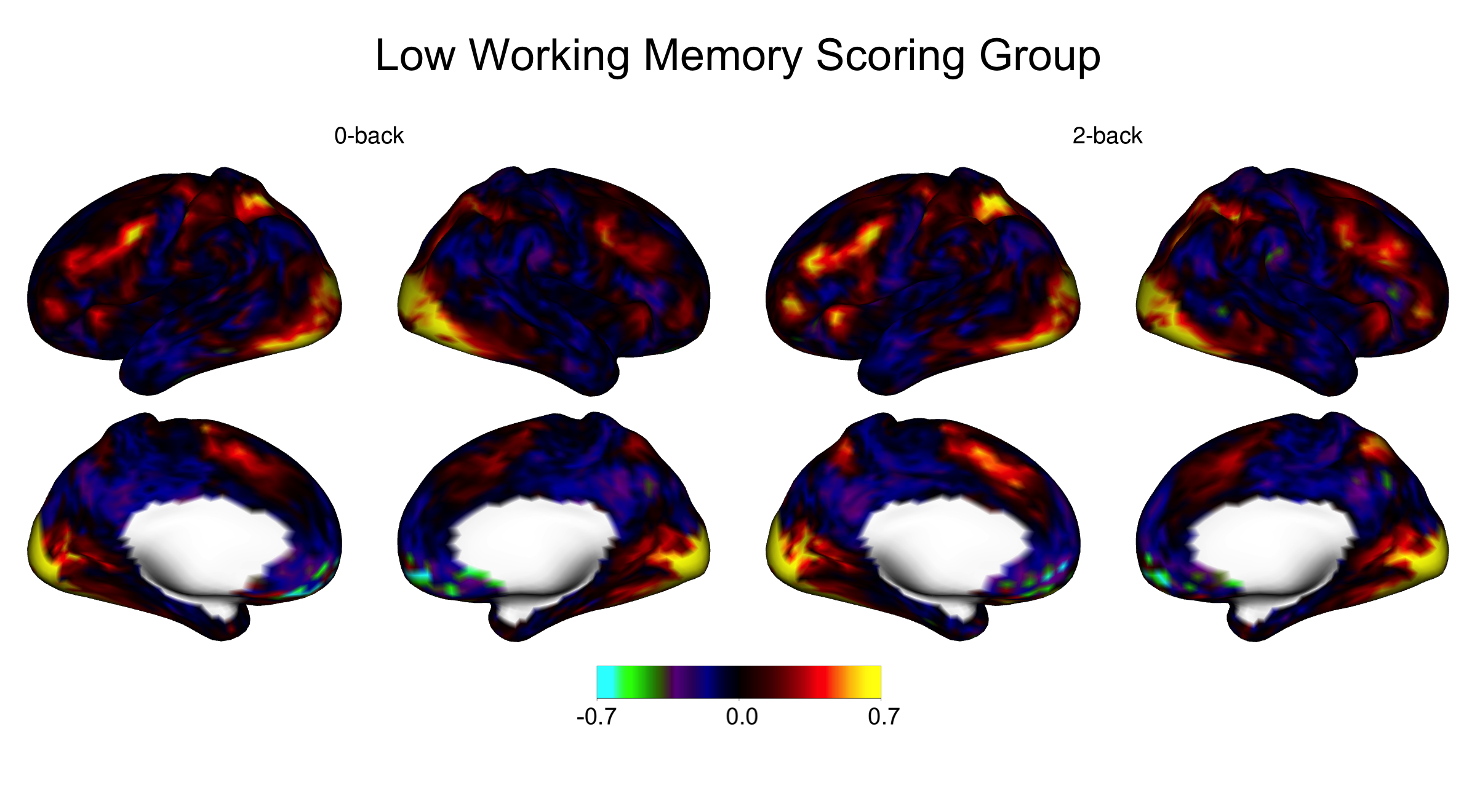}
        \caption{Activation field estimates for the `0-back' 
        (left) and the `2-back' 
        (right) tasks, averaged across sessions and subjects within
        the high (top) and low (bottom) working memory scoring groups.}
        \label{fig:application_beta_estimates}
    \end{figure}

Figure~\ref{fig:application_beta_estimates} 
shows the activation field estimates for the `0-back' task (left) and `2-back' task (right), taken as
averages across sessions and subjects belonging to the same group (high working memory score, top, and low working memory score, bottom). Because \emph{NSVR-BayesGLM}
allows non-stationarity, we can capture activation patterns 
with noticeably different degrees of smoothness across
regions. In both tasks, 
activation is, as expected,
most prominent in the striate and extrastriate visual cortex (BA V1, BA V2), followed by subregions of several functional divisions such as precentral ventral attention (BA 6, BA 44), salience/ventral attention (BA 44, BA 6), sommatosensory motor (BA 1, BA 2, BA 4p), prefrontal cortex (BA 45),  intraparietal sulcus (IPS), and superior parietal lobule (SPL). This is consistent with the conclusion in a recent meta analysis of visual working memory task
\citep{li22}. Notably, the IPS and SPL have been documented to play a crucial role in visuo-spatial attention and the analysis of spatial elements \citep[see, e.g.,][]{papadopoulos18, macsweeney08}. Functional regions that contain a larger, and/or more significantly activated areas during the `2-back' 
task include the precentral dorsal attention, salience/ventral attention, prefrontal cortex, inferior parietal lobule and precuneus. While the prefrontal cortex is 
key 
in regulating a large  number of higher-order executive functions, including working memory, the precuneus is well-known for its role in visuo-spatial imagery and episodic memory retrieval \citep[see, e.g.,][]{cavanna06}. When
contrasting high-
and low-scoring 
groups, 
the same regions seem to be activated in each task, but interestingly
activation magnitudes are higher across the board in the low scoring group. This
may suggest an increased effort in performing the tasks from subjects
who score lower in the working memory test. 


\begin{figure}[!b]
        \centering
         \includegraphics[width=\textwidth]{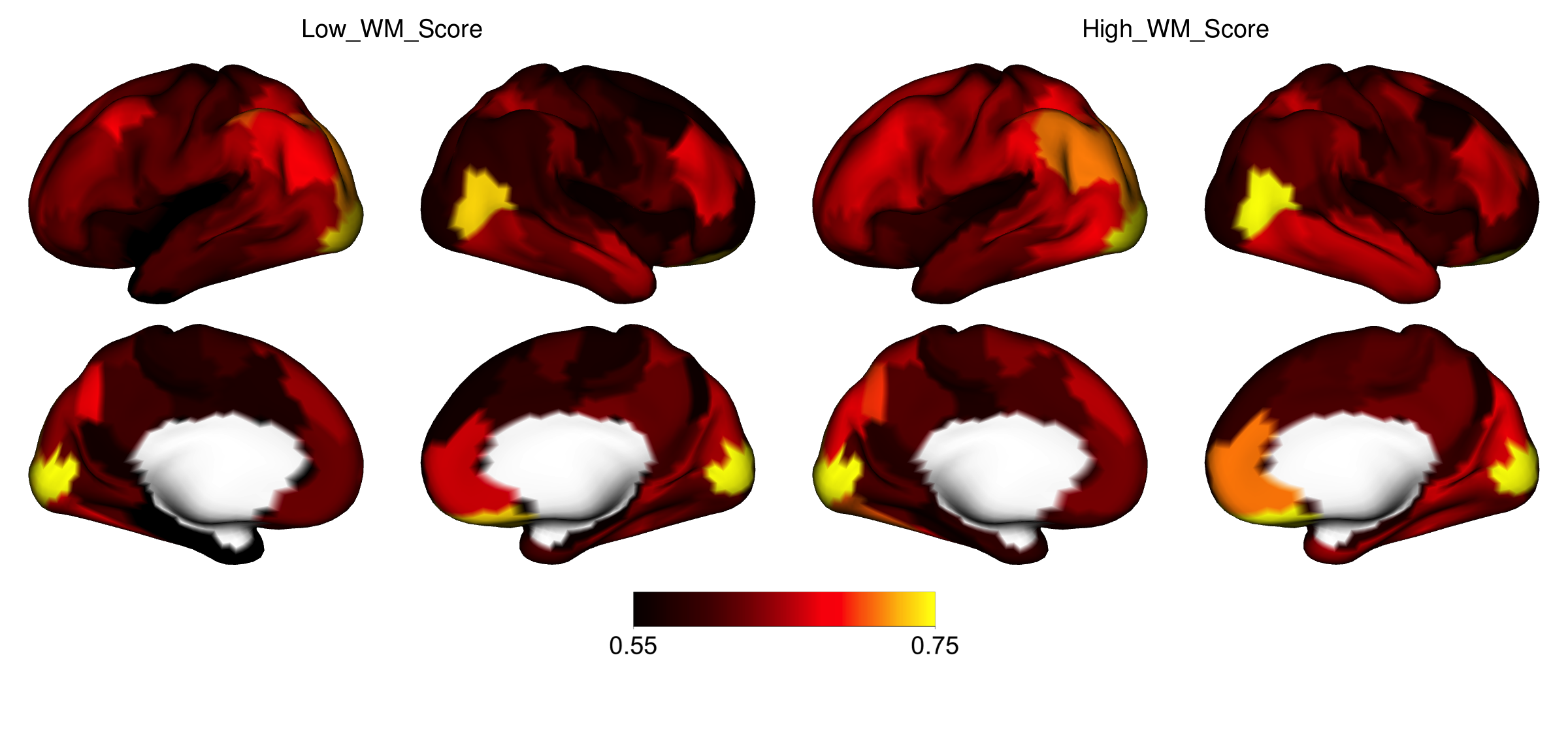}
        \caption{Hurst parameter estimates, averaged across subjects within the  low (left) and high (right) working memory scoring groups.
        No regions exhibit short-range dependence in either group, and high-scoring subjects exhibit longer range dependences than low-scoring subjects.}
        \label{fig:application_result_H}
\end{figure}
    
Figure~\ref{fig:application_result_H} 
shows Hurst parameter estimates for different brain regions, taken again as averages across subjects belonging to the same group. We note that using $n_H = 5$, all regions report Hurst estimates larger than $0.5$, 
which strengthen the case for an approach that, like \emph{NSVR-BayesGLM}, can accommodate long range dependences. Interestingly, subjects who score high 
in the working memory test
have noticeably higher Hurst estimates across almost all brain regions. 
This suggest an intriguing association between temporal dependence ranges and working memory performance -- and one that our region-specific Hurst estimates, unlike the single score value of the test, allow us to articulate spatially.
While this is only a preliminary result, it may point towards a promising, dependence range-based approach to the study of 
memory-related 
diseases. 
It also further strengthens our recommendation against prewhitening, which erases potentially valuable information contained in model residuals.

\section{Discussion}
\label{discussion}
In this article
we propose \emph{NSVR-BayesGLM}, a first-of-its-kind approach that fully integrates the modelling of non-stationary spatial dependence and varying ranges of temporal dependence in cs-fMRI data. The former is implemented 
through an extension of the spatial SPDE prior approach in \citet{mejia20}, which 
allows for spatial non-stationarity 
driven by local spatial features of the data 
and results in a sparsely parameterized model. Different ranges of temporal dependence are also modeled in a data-driven fashion,
combining the use of several fractional Gaussion processes of varying persistence and wavelet
transformations. We demonstrate via simulations that \emph{NSVR-BayesGLM} 
can better accommodate local activation features
compared to a stationary counterpart, while retaining comparable power and improving false positive control. 
Notably, in addition to assuming stationarity, the competing method employs autoregressive 
models of order $p = 6$ to prewhiten the data, while \emph{NSVR-BayesGLM}
accurately estimates and incorporates different ranges of temporal dependence at different locations. Finally, we apply
\emph{NSVR-BayesGLM} 
to a visual working memory cs-fMRI data set. The activation patterns are found to be consistent with existing literature. Moreover, \emph{NSVR-BayesGLM} allows us to unveil different degrees of smoothness in the signals characterizing different regions, and an intriguing association between 
estimated temporal dependence profiles and working memory performance -- something that may point towards further applications in the study of memory-related diseases. 

A limitation of the current 
proposal is that, while advancing the modeling of spatial and temporal dependence with respect to existing approaches, it still treats them as separate.
Spatial dependence is embedded within the activation coefficients' prior hyper-parameters, while temporal dependence is imposed upon the error terms and is reflected directly in the likelihood. In reality, 
it is likely that the spatial correlations 
characterizing activation fields change with time, e.g., as task stimuli switch on and off. 
Ideally, one should explicitly incorporate 
interactions between spatial and temporal dependence -- but pursuing such an extension would still be prohibitive in terms of computational burden.  

Indeed, 
computational cost remains a major challenge that 
influences modeling choices. As reported in Section~\ref{sec:application}, the current model estimation 
has a runtime of approximately 30 minutes per hemisphere, compared to \emph{BayesGLM}'s 45 seconds. The main
burden comes from inefficient memory allocation for large spatio-temporal models;
\textsf{R-INLA} takes up nearly 85GB of RAM during the estimation stage for a data set of $\approx 3000$ locations/vertices and $\approx 500$ time points. Note that, if we use prewhitening and do not explicitly model temporal dependence (as in the case of \emph{BayesGLM}), the non-stationary spatial model is just as fast and memory-efficient.
For the purpose of circumventing memory issue and reducing computational burden, we 
could consider an approach that combines INLA with the MCMC algorithm. 
This works by iterating between drawing MCMC samples from the posterior marginals of temporal dependence hyperparameters, and fitting only the non-stationary spatial component with INLA, conditioning on the MCMC samples. This ``fusion'' approach, 
also known as the Metropolis-Hastings with INLA, was proposed by \citet{rubio17}
and shown to have a performance 
comparable to INLA via numerical examples and heuristic arguments, for some selected models. For our model, potential complications with parameter support boundaries and approximation errors of the conditional marginal likelihood might exist, and will need to be studied in future work. 
Complete code 
implementing the current version of \emph{NSVR-BayesGLM} as well as data for reproducing our simulations are available at \url{https://github.com/hqd1/SPfmri}.


\section*{Acknowledgments}
We utilized data 
from the Human Connectome Project, WU-Minn Consortium (Principal Investigators: David Van Essen and Kamil Ugurbil; 1U54MH091657) funded by the 16 NIH Institutes and Centers that support the NIH Blueprint for Neuroscience Research,
and by the McDonnell Center for Systems Neuroscience at Washington University. We utilized substantial portions of the \textsf{R} code developed in 
\citet{mejia20}, we thank the authors for making such code available and for providing advice. 
The work of M.A.~Cremona was partially supported by the Natural Sciences and Engineering Research Council of Canada (NSERC), by the Fonds de recherche du Québec Health (FRQS), and by the Faculty of Business Administration, Université Laval.  The work of 
F.~Chiaromonte was partially supported by the Huck Institute of the Life Sciences of the Pennsylvania State University. \\
M.A. Cremona is also affiliated with CHU de Québec -- Université Laval Research Center, Canada. F. Chiaromonte is also affiliated with the Sant'Anna School of Advanced Studies, Italy.
\bibliographystyle{agsm}
\bibliography{trial}
\end{document}